\documentclass[preprint,prb,aps,showpacs,showkeys,preprintnumbers,amsmath,amssymb]{revtex4}

\usepackage{graphicx}
\usepackage{epsfig}
\usepackage{bm}
\usepackage[active]{srcltx}
\usepackage[latin1,applemac]{inputenc}
\usepackage{natbib}
\usepackage{amsfonts}
\begin{document}

\title{	Joule heating and high frequency nonlinear effects in the surface impedance of high 
T$_{c}$ superconductors }

\author{	Julien Kermorvant,$^{1}$ Cornelis Jacominus van der Beek,$^{2}$  Jean-Claude Mage$^{1}$ , Bruno 
Marcilhac,$^{1}$  Yves Lema\^{i}tre,$^{1}$  Javier Briatico,$^{1}$  Rozenn Bernard,$^{1}$  Javier Villegas$^{1}$ }
\affiliation{$^{1}$ Unit\'{e} Mixte de Recherche en Physique UMR 137 CNRS-THALES}

\affiliation{$^{2}$ Laboratoire des Solides Irradi\'{e}s, CNRS-UMR 7642 \& CEA-DSM-IRAMIS, Ecole Polytechnique, F 91128 Palaiseau cedex, France}

\begin{abstract}

Using the dielectric resonator method, we have investigated 
nonlinearities in the surface impedance $Z_{s}=R_{s}+jX_{s}$ of 
YBa$_{2}$Cu$_{3}$O$_{7-\delta}$ thin films at 10 GHz as function of 
the incident microwave power level and temperature. The use of a 
rutile dielectric resonator allows us to measure the precise 
temperature of the films. We conclusively show that the usually 
observed increase of the surface resistance of 
YBa$_{2}$Cu$_{3}$O$_{7-\delta}$ thin film as function of microwave 
power is due to local heating.
\end{abstract}


\maketitle

\section{Introduction}
\label{section:Introduction}
High Temperature Superconductor (HTSC) films are suitable candidates 
for the improvement of microwave receiver performance  because of 
their low surface resistance.\cite{Hein96,Gallop97,Portis93,Lancaster97} The surface impedance of HTSC 
materials presents a strong dependence on the magnitude of the 
incident microwave magnetic field, $H_{rf}$. Typically, a nonlinear 
behavior is observed above a certain value of $H_{rf}$. This 
nonlinearity leads to unacceptable microwave power losses. Microwave 
losses are characterized by a decrease of the quality factor $Q$ of 
superconducting resonators and filters and a downward shift of the 
resonant frequency of the former.\cite{Samoilova95,Schen94,Hein97} The surface impedance of HTSC 
has been studied by many groups, using either the dielectric 
resonator technique or the stripline resonator technique. In spite 
of numerous experimental studies the physical origin of the observed 
nonlinearities is still under debate and the subject of present-day 
experimental investigation.\cite{Lee97,Wosik99,Nguyen2000,Hein2000,Xin99}

It has been proposed that a simple way to differentiate among the 
mechanisms of nonlinearity of the surface impedance is the 
examination of the $r$ parameter.\cite{Herd97} This quantity is defined as 
the ratio between the microwave field dependence of the surface 
reactance $\Delta X_S(H_{rf})$  and of the surface resistance $\Delta 
R_S(H_{rf})$,

\begin{equation}
  r=\dfrac{\Delta X_S(H_{rf})}{\Delta R_S(H_{rf})}.  
\end{equation}

Table 1 gives an overview of possible mechanisms leading to the 
nonlinear behavior of the surface 
impedance.\cite{Lahl2004,Oates2002,Golosovsky98} First is the
intrinsic nonlinearity due to pair-breaking. The nonlinearity is then 
related to the increase of the quasi-particle density $n_{qp}$ 
\cite{Dahm96,Dahm97} when the $H_{rf}$-induced current density is of the order of 
magnitude of the pair-breaking current density. A nonlinearity based 
on this effect has been predicted and investigated using a 
phenomenological expression for a nonlinear penetration depth. If the 
nonlinearity is dominated by this intrinsic mechanism, the 
$r$-parameter should be small and strongly frequency-dependent. This 
differs from the experimentally observed nonlinearities. Hysteretic 
losses are also proposed to be 
signifiant.\cite{Hylton98,Hylton89,Dimas90} The weakly 
coupled-grain model holds that the large surface resistance of highly 
granular high-T$_{c}$ superconductors as compared to single crystals can 
be explained by the different morphology. The polycrystaline samples 
can be modeled as a network of Josephson junctions. Nonlinear behavior 
is expected at rf-current densities that are limited by critical 
current density of the constituant Josephson junctions.  This model 
yields a very small $r$-parameter with a strong dependence on 
temperature and frequency. In the case of granular films the coupled-grain
 model describes the microwave nonlinearities fairly well. 
Nevertheless, it fails to describe strong nonlinearities in epitaxial 
films. Vortex penetration and creep into grain boundaries and/or 
weak links are also proposed as a possible source of microwave 
hysteretic losses. Vortex generation by the microwave magnetic field 
has been predicted \cite{Dahm99,Sridhar94,McDonald97} and the Bean model has been extented to 
account for microwave nonlinearity in HTSC films. The dependence of 
$R_{s}$ on $X_{s}$ is almost linear (with slope $r$). Experimental 
values are close to those predicted by the model.
The last possible effect to explain nonlinearity is local or uniform 
heating.\cite{Hein99,Wosik97,Pukhov97,Pukhov97ii} It has been proposed that heating can occur in 
superconducting thin films. This effect appears in the microwave 
frequency range,  particularly in continuous mode, but also in pulsed 
mode, as function of the pulse period. Heating is significant above 
a certain value of the incident microwave power, and causes the 
transition to the normal state of weaker superconducting regions such 
as weak links or local defects. Heating and heat transfer to the 
substrate are shown to play an important role.

We have used the dielectric resonator method  in order to measure the 
surface impedance  of YBa$_{2}$Cu$_{3}$O$_{7-\delta}$ thin films from 
various sources.
We present a study of both the temperature and the microwave power 
level dependence. The use of a rutile dielectric resonator allows us to 
measure the precise temperature of the films. We show that the 
usually observed increase of the surface resistance of 
YBa$_{2}$Cu$_{3}$O$_{7-\delta}$ thin films as a function of microwave 
power is due to local heating.

\section{Experimental details}
\label{section:Exp}

Measurements of the surface impedance $Z_ {s}=R_{s}+jX_{s}$ were 
performed on a series of YBa$_{2}$Cu$_{3}$O$_{7-\delta}$ thin films, 
denoted SY211 and obtained at THALES by inverted cylindical hollow 
cathode dc sputtering. A second series, labelled TM MgO and TM LAO, 
was acquired from THEVA Inc..\cite{THEVA} This series was prepared by 
reactive thermal evaporation. $Z_{s}$ was measured by the dielectric 
resonator method.\cite{Hein97ii,Marzierska98,Wilker2003,Lee2005,Diete97} The film thicknesses and the critical 
temperatures as measured by ac-suceptibility are gathered in Table 
2. Surface morphology imaging is shown in Fig. 1 for the three 
studied samples. X-ray diffraction, using a Bragg-Brentano diffractometer, showed all films to be epitaxial with the [001] orientation.

For the surface impedance measurements, we have used a cylindrically 
shaped rutile resonator of height 1 mm and diameter 7 mm, which is 
directly placed onto the sample. The resonant frequency in the 
$TE_{011}$ mode is near 10 GHz. Rutile is well-known for its  small 
tangent loss ( $\tan\delta=10^{-5}$ at 77 K, 10 GHz) and its very 
high dielectric constant ($\epsilon$ = 105 at 77 K).\cite{Tobar97,Mage90} The 
resonator is excited by an adjustable coupling loop; the distance 
between the resonator and the loop is controlled in order to maintain 
critical coupling during the experiment. The whole assembly is placed 
inside an oxygen-free high conductivity copper cavity, mounted onto a 
cryocooler cold head. Temperature stability is better than 1 mK over 
the range 30 K-90 K.

 For each sample, we measure the resonant frequency $f_{0}$ and the 
loaded $Q$-factor of the fundamental resonance of the resonator. At 
each microwave input power level, the reflection coefficient from the 
resonator, or $S_{11}$ parameter, is measured with a network analyser 
in continuous mode, coupled to a microwave amplifier for the high 
power regime. This experimental set-up allows us to measure the 
surface impedance over a input microwave power ranging from 0.01 mW 
to 100 mW. 

The loaded $Q$-factor of the resonator is given by:

\begin{equation}   Q_{L}=\dfrac{f_{0}}{\Delta f},   \end{equation}

\noindent where $f_{0}$ and $\Delta f$ are, respectively, the 
resonant frequency and the -3 dB bandwidth in log scale corresponding 
to the resonant peak, see Fig. 2.

The unloaded $Q$-factor is defined by :

\begin{equation}  
 Q_{0}=(1+\beta)Q_{L},  
\end{equation}

\noindent with $\beta$ the coupling constant. All measurements were 
performed under critical coupling {\it i.e} $\beta=1$, and the 
unloaded $Q$-factor $Q_{0}=2Q_{L}$.
The inverse $Q_{0}^{-1}$ is the sum of different contributions due to 
the resonator itself, the copper cavity,  and the 
YBa$_{2}$Cu$_{3}$O$_{7-\delta}$ thin film, such that 

\begin{equation}  
%\dfrac{1}{Q_{0}}=\dfrac{1}{Q_{resonator}}+\dfrac{1}{Q_{YBCO}}+\dfrac{1}{Q_{Cu}}.
\frac{1}{Q_{0}}=\frac{1}{Q_{resonator}}+\frac{1}{Q_{YBCO}}+\frac{1}{Q_{Cu}}.    
\end{equation}

\noindent Here ${Q^{-1}_{resonator}}=A\tan\delta_{TiO_{2}}$ is due to 
the dielectric losses, ${Q^{-1}_{Cu}}=CR_{s,Cu}$ to the microwave 
losses in the copper, and  ${Q^{-1}_{YBCO}}=B R_{s,YBCO}$ arises from 
the microwave losses in the YBa$_{2}$Cu$_{3}$O$_{7-\delta}$ film. The 
geometrical factors A=0.9871, B=1.75.$10^{-2}$, C=3.665.$10^{-5}$ are 
calculated using a numerical simulation (HFSS software 
\cite{HFSS}). 
    
    The surface resistance is obtained as 

\begin{equation}
     R_{s,YBCO}=\dfrac{1}{B}\left( 
\dfrac{1}{Q_{0}}-A\tan\delta_{TiO_{2}}-CR_{s,Cu}\right). 
  \end{equation}

\section{Results}
\label{section:Results}

In order to understand the variation of the $Q$-factor and the 
resonant frequency with increasing microwave power, we have measured 
the temperature dependence of the resonator's properties. Fig. 3(a) 
represents the temperature dependence of the TiO$_{2}$ resonator resonant 
frequency in the limit of small microwave power $P_{rf}$, for three 
different configurations. In the first configuration, the TiO$_{2}$ 
resonator is directly placed on the copper cavity; in the second, the 
resonator is placed on an MgO substrate; finally the resonator is 
placed on the YBa$_{2}$Cu$_{3}$O$_{7-\delta}$ film, itself deposited 
on MgO. The absolute value of the resonant frequency, $f_{0}$, 
depends on the distance between the resonator and the conducting wall 
of the copper cavity or of the superconducting layer. The presence of 
dielectric MgO leads to a lower resonant frequency than the presence 
of the conducting YBa$_{2}$Cu$_{3}$O$_{7-\delta}$ layer. However, the 
temperature dependence of the frequency shift of the rutile resonator 
does not depend on the nature of the support. This shows that the 
thermal conductivity between the cryocooler cold-head and the rutile 
resonator is not significantly affected by the intercalation of the 
500 $\mu$m-thick MgO and the 400 $n$m-thick superconducting layer.

Fig. 3(b) represents the temperature dependence of the resonant 
frequency shift of the rutile resonator and of a MgO resonator with 
$f_{0}$ near 8 GHz. Clearly, the variation with temperature of the MgO 
resonant frequency is much weaker than that of rutile. The 
temperature dependence of the resonant frequency is the direct 
consequence of the increase (resp. decrease) with temperature of the 
dielectric constant $\epsilon(T)$ of rutile (resp. MgO).

In Fig. 4 (a,b) we present the surface resistance and the resonant 
frequency in the presence of the investigated YBa$_{2}$Cu$_{3}$O$_{7-\delta}$ films 
at a given temperature of 74 K, as function of an effective microwave 
reactive power. This is defined by $\tilde{P}_{rf}=P_{incident,rf} \times 
Q_{L}$, a parameter introduced to quantitatively compare data obtained 
on the different films. Namely, each different film leads to a different 
values of $Q_{L}$ and absorbed power for the same incident microwave 
power.  The zero power limit is taken as those values of $\tilde{P}_{rf}$ below 
which $R_{s}(\tilde{P}_{rf})$ is essentially $\tilde{P}_{rf}$-independent.
Curves for different films present the same behavior, {\it i.e} 
$R_{s}(\tilde{P}_{rf})$ and $f_{0}(\tilde{P}_{rf})$ are independent of the microwave 
field in the zero field limit and become nonlinear (increase rapidly) 
above a threshold value $\tilde{P}_{rf}$.

 Contrary to what is expected and usually observed by using sapphire 
or MgO resonators, the resonant frequency $f_{0}$ of the rutile 
resonator also increases with increasing microwave power.
 We ascertain that the increase of $f_{0}(\tilde{P}_{rf})$ is due to the 
heating of the rutile resonator by the 
YBa$_{2}$Cu$_{3}$O$_{7-\delta}$ film.
In order to demonstrate this effect, we have also measured the 
temperature dependence of the surface resistance and resonant 
frequency in the limit of small $\tilde{P}_{rf}$. The surface resistance at 
$\tilde{P}_{rf} \rightarrow 0$  of the YBa$_{2}$Cu$_{3}$O$_{7-\delta}$ films, 
shown by the open symbols in Fig. 5(a),  shows the usual monotonous 
increase with temperature for all samples. Concerning the temperature 
dependence of the resonant frequency, we observe an increase  with 
temperature, as discussed previously, Fig.5(b). The only observed 
difference is the nearly constant frequency offset between the three 
curves. 

By fitting the $T(f_{0})$ curve of  Fig. 5(b) and substituting the interpolated values for the $f_{0}$-values measured in the swept-power experiment depicted in Fig. 4(b), 
we estimate the temperature variation $T(\tilde{P}_{rf})$ of the resonator in the 
latter. Fig. 6 shows the temperature variation of the
resonator obtained in this way, at nominal 
temperatures of 63 K (a) and 74 K (b). Fig. 7 shows the normalized 
resonant frequency $f_{0}$ of the rutile resonator placed on a 
superconducting layer (black square) and on a MgO substrate 
(black star) as a function of the microwave reactive power. This graph
shows that the contribution to the observed nonlinearities arising from the losses in the dielectric material is negligible compared to
the losses in the superconductor.

We now compare the $R_{S}$ data obtained as function of the temperature increase due to increasing dissipation in the swept-power experiments and the low-power $R_{S}(T)$
data. Fig. 5(a) shows a superposition of $R_{S}(T)$ obtained from swept-power experiments at nominal temperatures of 63 K and 74 K (Fig. 6 closed symbols) and the low-power
$R_{S}(T)$ data (open symbols).The data are perfectly superimposed, this means that no intrinsic 
$\tilde{P}_{rf}$-dependence of the surface resistance is measured, and that 
any observed nonlinearity is the consequence of Joule heating.

\section{Discussion}
\label{section:Discussion}

The temperature dependence of the resonant frequency is the direct 
consequence of the increase with temperature of the dielectric 
constant $\epsilon(T)$ of rutile. Note that its behavior is opposite 
to the decrease with temperature of the dielectric constant of more 
commonly used sapphire or MgO resonators. Moreover, the variation 
with temperature of the MgO resonant frequency is much weaker than 
that of rutile. By consequence, it is difficult to separate the 
evolution of the intrinsic change of a MgO or sapphire resonator's 
frequency from that caused by the temperature variation of a 
superposed superconducting film: both weakly decrease as function of 
temperature. However, the intrinsic evolution of the rutile 
resonator's frequency is opposite to that expected from the presence 
of the superconducting film. A measurement of the rutile's resonator 
frequency can thus unambiguously serve as a local temperature 
measurement.

In general, superconducting thin films contain resistive defects such 
as conducting precipitates and weak-links of Josephson junctions. In 
their experiments, Obara et al. \cite{Obara2006} show a correlation between the 
$H_{rf}$-dependence of the surface resistance and the value of the dc 
critical current density. Their interpretation is these correlations 
indicate that the power dependence of $R_{s}$ is due to the intrinsic 
properties of the films, such as pair-breaking. However, nonlinearities 
usually occur at much lower $H_{rf}$ fields than those at which the 
intrinsic nonlinearities are expected.\cite{Nguyen93,Keskin99,Gaganidze2000,Halbriter95} Halbriter et al. 
\cite{Halbriter2001} show that at very low microwave fields, nonlinearities can be 
explained by Josephson fluxon penetration along weak links; at higher 
fields flux flow losses may also participate.\cite{Gurevich93} Wosik et al. 
insist on the importance of thermal effects as the root of the 
nonlinear behavior.\cite{Wosik97ii,Wosik96,Wosik99ii} They have used the pulsed measurement 
method, which, in principle, prevents or at least reduces heating of 
the films. These experiments were repeated using  thermally isolated 
films as well as films that are thermally connected through the 
substrate with the heat sink cooled at 20 K. The 
authors\cite{Wosik97ii,Wosik96,Wosik99ii} have 
shown that the temperature increase of an isolated sample can reach 
30 K. For the case of a film thermally connected to the heat sink the 
temperature rise will be of only 2.5 K.\\

\section{Concluding remarks}
\label{Conclude}

In our experiment we have clearly shown that the use of the rutile 
dielectric material is a good way for a direct measurement of the 
temperature of superconducting thin films, because of the large 
temperature dependence of the dielectric constant. The behavior of the resonant frequency, opposite 
to that of the commonly used sapphire or MgO 
resonators, indicates that the resonant frequency shift 
observed as function of  the applied microwave power is mainly due to 
heating of superconducting films. The perfect superposition of the temperature-dependent surface resistance obtained from 
swept-power experiments with the temperature dependence of $R_{s}$ measured in the limit of low microwave power shows that intrinsic
nonlinearities of the superconducting films do not contribute to heating. Therefore, Joule heating must be due to either flux-flow losses
or to quasi-particle resistive losses. 

\pagebreak

\newpage

\begin{table}[h!]
\caption{Possible mechanisms causing nonlinearity of the surface impedance. }
\begin{tabular}{cccccc}
     \hline
     \hline
               
               \multicolumn{1}{|c|}{Mechanism}  &  \multicolumn{1}{|c|}{Ref} & \multicolumn{1}{|c|}{$R_{s}$ and $X_{s}$ microwave} &  \multicolumn{1}{|c|}{$r$ value}&       \multicolumn{1}{|c|}{ Temperature}     &      \multicolumn{1}{|c|}{Frequency}  \\ 
                                      
   \multicolumn{1}{|c|}{}            &      \multicolumn{1}{|c|}{}    & \multicolumn{1}{|c|}{ field dependence }     &          \multicolumn{1}{|c|}{}      & \multicolumn{1}{|c|}{dependence of $r$} &   \multicolumn{1}{|c|}{dependence of $r$}                                              
             
\cr \hline

             Intrinsic non linearity & [17-18] &$ \propto H^{2}_{rf}$ low power &   $10^{-2}$ & Increase with $T$ & $\propto \omega$ \\
                                                  
              Pairbreaking           &       &  $\propto H^{4}_{rf}$ high power&                     &                               &

\cr
   
           Weakly coupled grain & [19-21]   &  $\propto H^{2}_{rf}$ & $ \>10^{-3}$& increase with $T$ & $\propto \omega$

\cr 

          Vortices in weak link & [22-24]  &    $\propto H_{rf}$ & $\leqq 1$ & $T$ independent & $\omega$ independent

\cr 

          Vortex penetration & [22-24]  &  $\propto H^{n}_{rf}$ $n\backsim4$ & const $\backsim 1$& $T$ independent & $\omega$ independent\\
        to the grains    &     &                                                                                  &                                      &                    &

\cr

        Uniform heating & [25-28]    &   $\propto H^{2}_{rf}$ & $ 10^{-2}$ & increase with $T$ & $\propto \omega$ 

\cr

       Heating of weak link & [25-28]    &  unknown &  $\backsimeq 1$& $T$ independent & $\omega$ independent 

\cr\hline
\hline
\end{tabular}

\end{table}

\newpage
\newpage
\begin{table}[h!]
\caption{Basic properties of the studied samples.}
\begin{tabular}{cccc}
\hline  
\hline

\multicolumn{1}{|c|}{ Name} &\multicolumn{1}{|c|}{ Substrate} &\multicolumn{1}{|c|}{ Thickness(nm)} & \multicolumn{1}{|c|}{T$_{c}$(K)}

\cr \hline 

SY211 & MgO & 400  & 88

\cr 

TM MgO & MgO & 700 & 88.5

\cr 

TM LAO &LaAlO$_{3}$&700 & 88.3

\cr\hline
\hline
\end{tabular}

\end{table}

\newpage 

\begin{table}[h!]
\caption{ $R_{s}$ and $f_{0}$ for $\tilde{P}_{rf}=P_{nl}$ at 74 K and in 
the limit of small power regime at $T=T_{nl}$.}
\begin{tabular}{ccccccc}

\hline
\hline
\multicolumn{1}{|c|}{name}&\multicolumn{1}{|c|}{ $\tilde{P}_{rf}$( W 
)}&\multicolumn{1}{|c|}{ $R_{s} (\tilde{P}_{rf}<P_{nl})$}&\multicolumn{1}{|c|}{ $f_{0}(\tilde{P}_{rf}<P_{nl})$} &\multicolumn{1}{|c|}{ $T_{nl}$(K)}& \multicolumn{1}{|c|}{$R_{s}(T<T_{nl})$}&\multicolumn{1}{|c|}{ $f_{0}(T<T_{nl})$}

\cr\hline

SY211&110.81 & 474.37 $\mu \Omega$&9.9805 GHz&55.12&176.43 $\mu \Omega$&9.8735 GHz

\cr 

TM LAO&135.26&302.34 $\mu \Omega$ &9.9758 GHz&63.74&173.72 $\mu \Omega$&9.9187 GHz

\cr

TM MgO&137.75&236.99 $\mu \Omega$&9.9740 GHz  &64.94&149.71 $\mu \Omega$&9.9284 GHz

\cr \hline
\hline
\end{tabular}

\end{table}

\newpage
\begin{figure}[h!]
\caption{Scanning electron microscope images for the 
three studied samples}
\end{figure}

\begin{figure}[h!]
\caption{ Plot of the resonant peak of the rutile resonator on 
YBa$_{2}$Cu$_{3}$O$_{7-\delta}$ film at different microwave input 
power levels.}
\end{figure}

\begin{figure}[h!]
\caption{(a) Plot of the resonant frequency of the 
TiO$_{2}$ resonator in three different configurations: Open squares: 
TiO$_{2}$ directly placed onto the copper cavity; closed stars: 
TiO$_{2}$ placed on an MgO substrate, itself placed in the copper cavity;  
open triangles TiO$_{2}$ placed on the superconducting layer deposited on 
MgO itself placed in the copper cavity. (b) Temperature dependence of the resonant frequency  
of the TiO$_{2}$ and MgO resonators.}
\end{figure}

\begin{figure}[h!]
\caption{ Dependence of the surface resistance (a) and of the resonant frequency 
(b) on microwave reactive power $\tilde{P}_{rf}$, at 74 K.}
\end{figure}

\begin{figure}[h!]
\caption{ Temperature dependence of the surface 
resistance (a). Open symbols denote the surface resistance as function of the directly measured 
temperature in the limit of small  microwave  power; closed 
symbols show the surface resistance measured in the swept-power experiment 
as a function of the calculated temperature. (b) The resonant frequency in the regime of low 
microwave power; the nearly constant offset between the three curves 
is due to the different values of the film and substrate thickness .}
\end{figure}

\begin{figure}[h!]
\caption{ Estimated temperature of the films, as deduced from the variation of the resonant frequency, as function of reactive 
microwave power, for a base (measurement) temperatures of (a) 63
K and (b) 74 K.}
\end{figure}

\begin{figure}[h!]
\caption{Resonant frequency of the rutile resonator as function of the microwave reactive power with the resonator placed on a superconducting layer (Black square) and on an MgO substrate (Black star)}
\end{figure}

\newpage
\begin{figure}[h!]
\includegraphics[width=8cm,keepaspectratio]{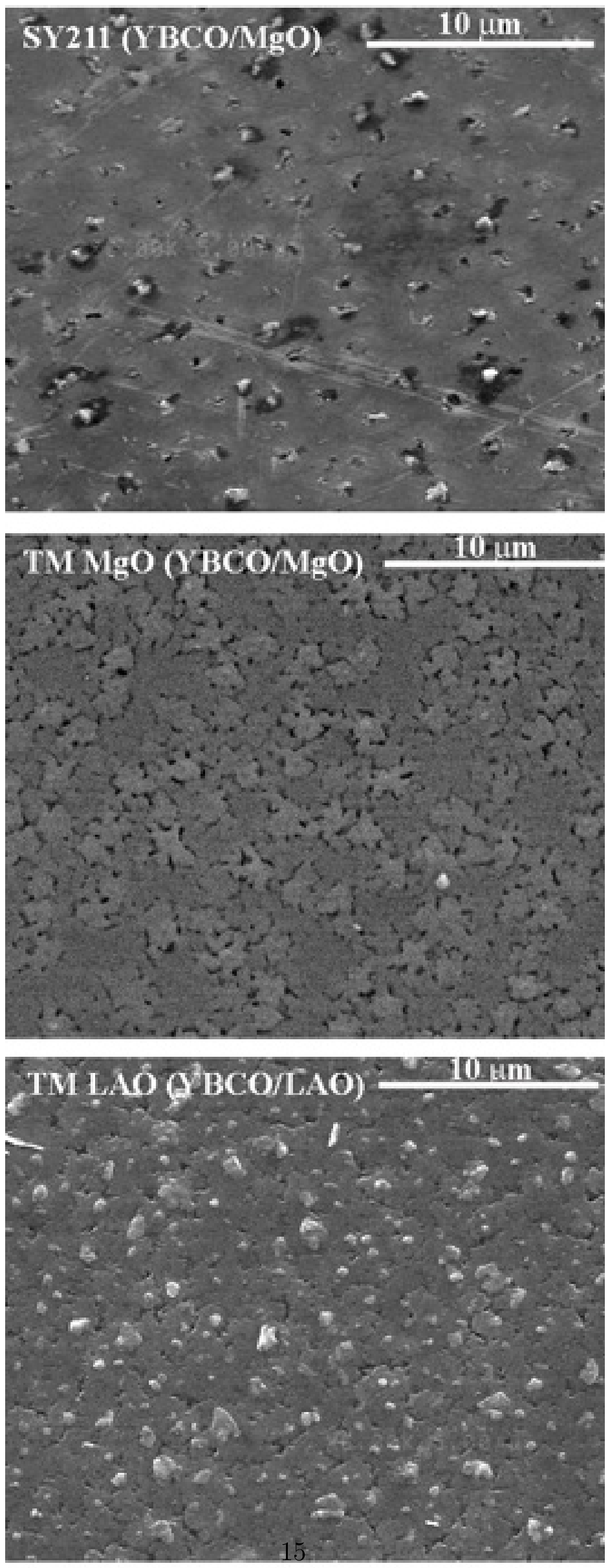} 
\begin{center}  Kermorvant et al. Fig. 1 \end{center}
\end{figure}

\newpage

\begin{figure}[h!]
\includegraphics[width=8cm,keepaspectratio]{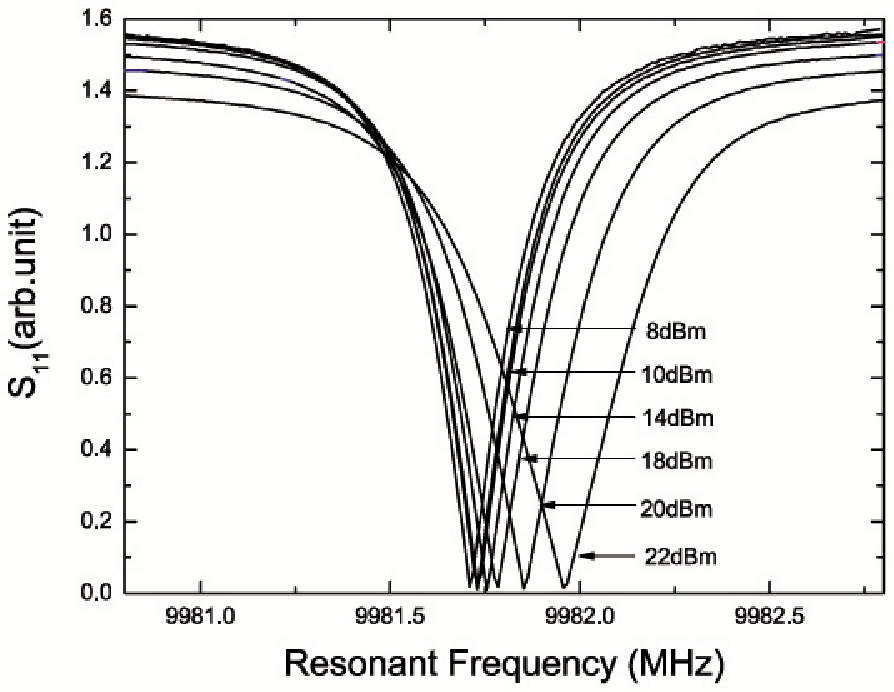} 
\begin{center}  Kermorvant et al. Fig. 2 \end{center}
\end{figure}

\newpage

\begin{figure}[h!]
\includegraphics[width=8cm,keepaspectratio]{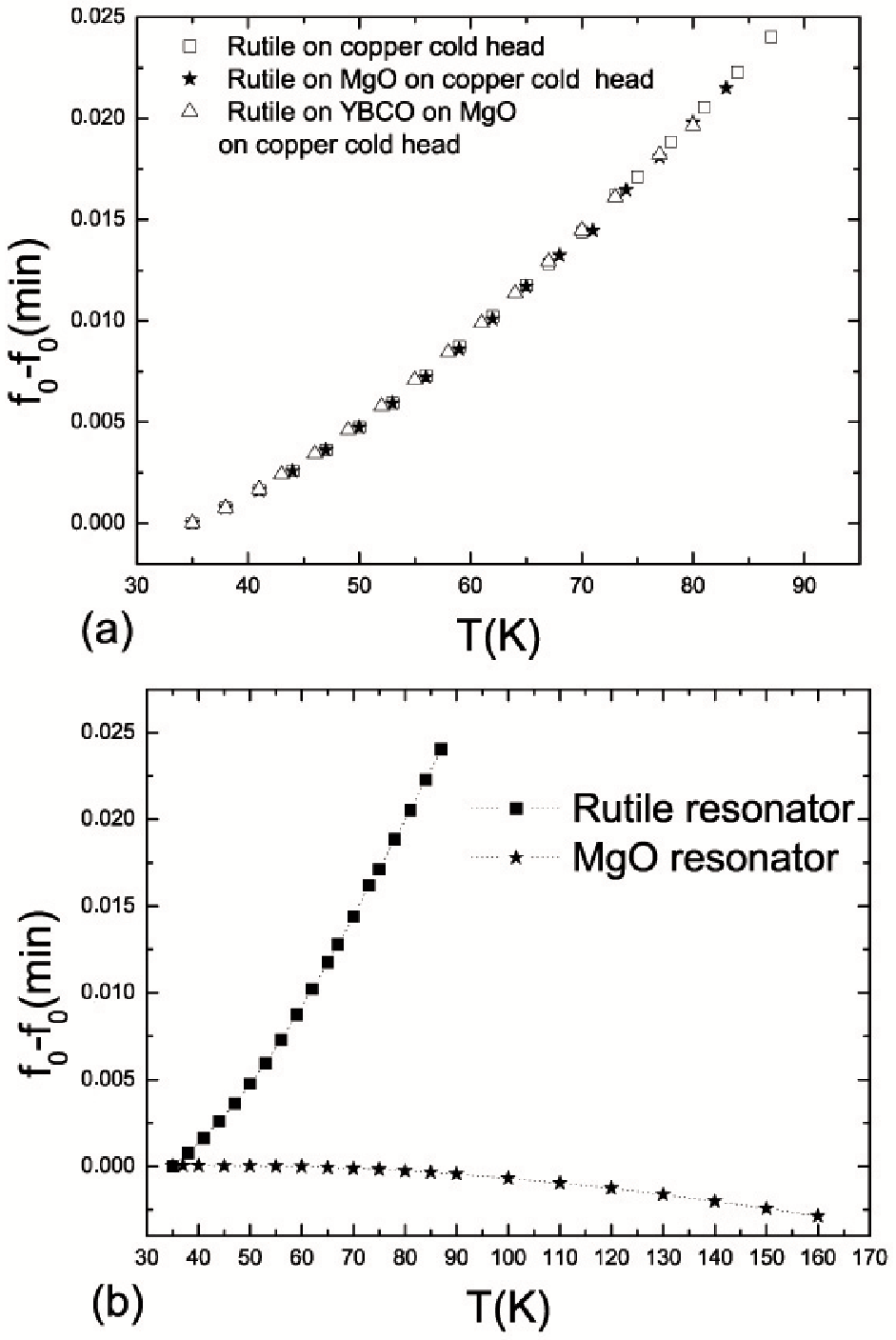} 
\begin{center}  Kermorvant et al. Fig. 3 \end{center}
\end{figure}

\newpage

\begin{figure}[h!]
\includegraphics[width=8cm,keepaspectratio]{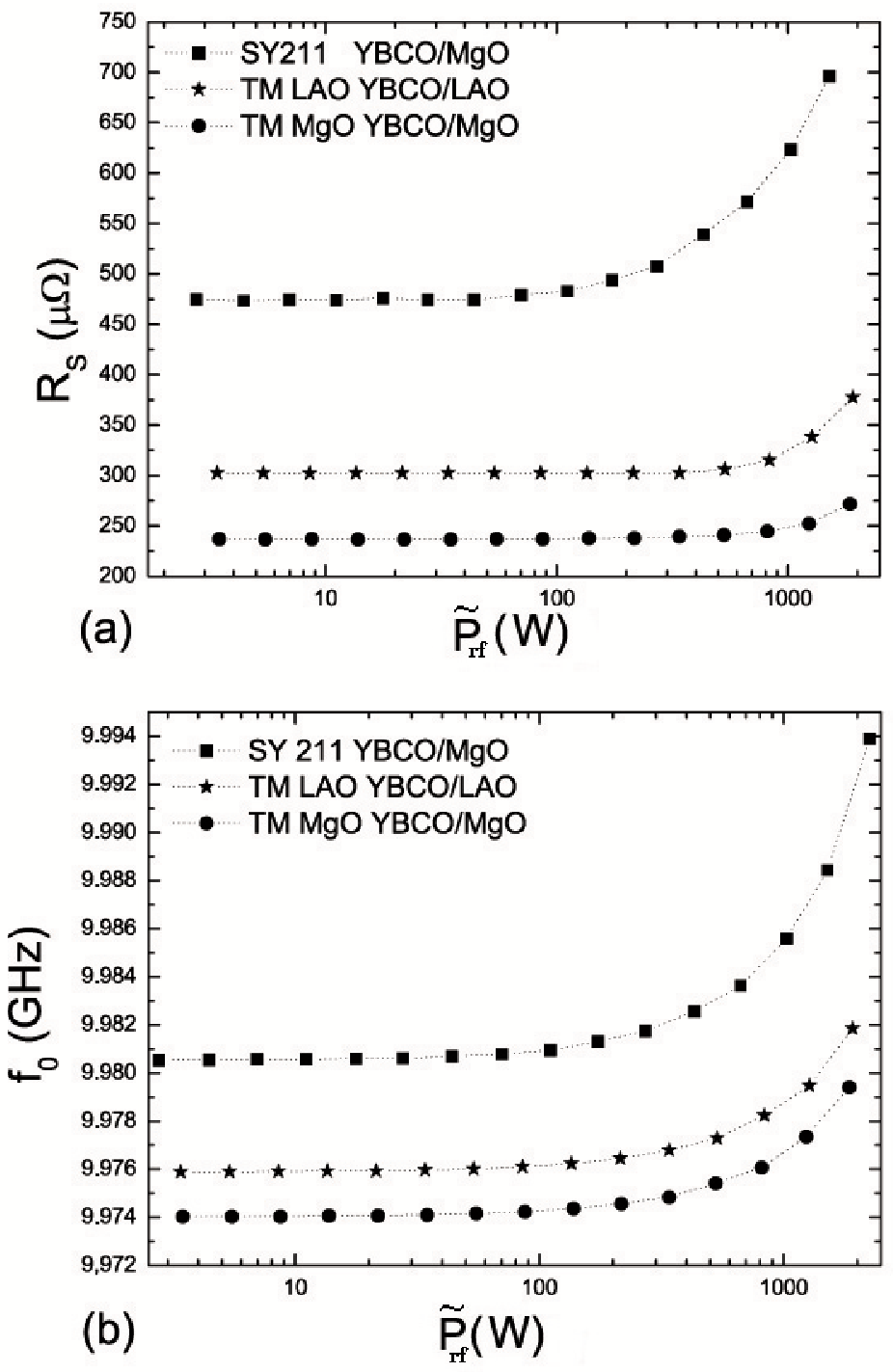} 
\begin{center}  Kermorvant et al. Fig. 4 \end{center}
\end{figure}

\pagebreak

\begin{figure}[h!]
\includegraphics[width=8cm,keepaspectratio]{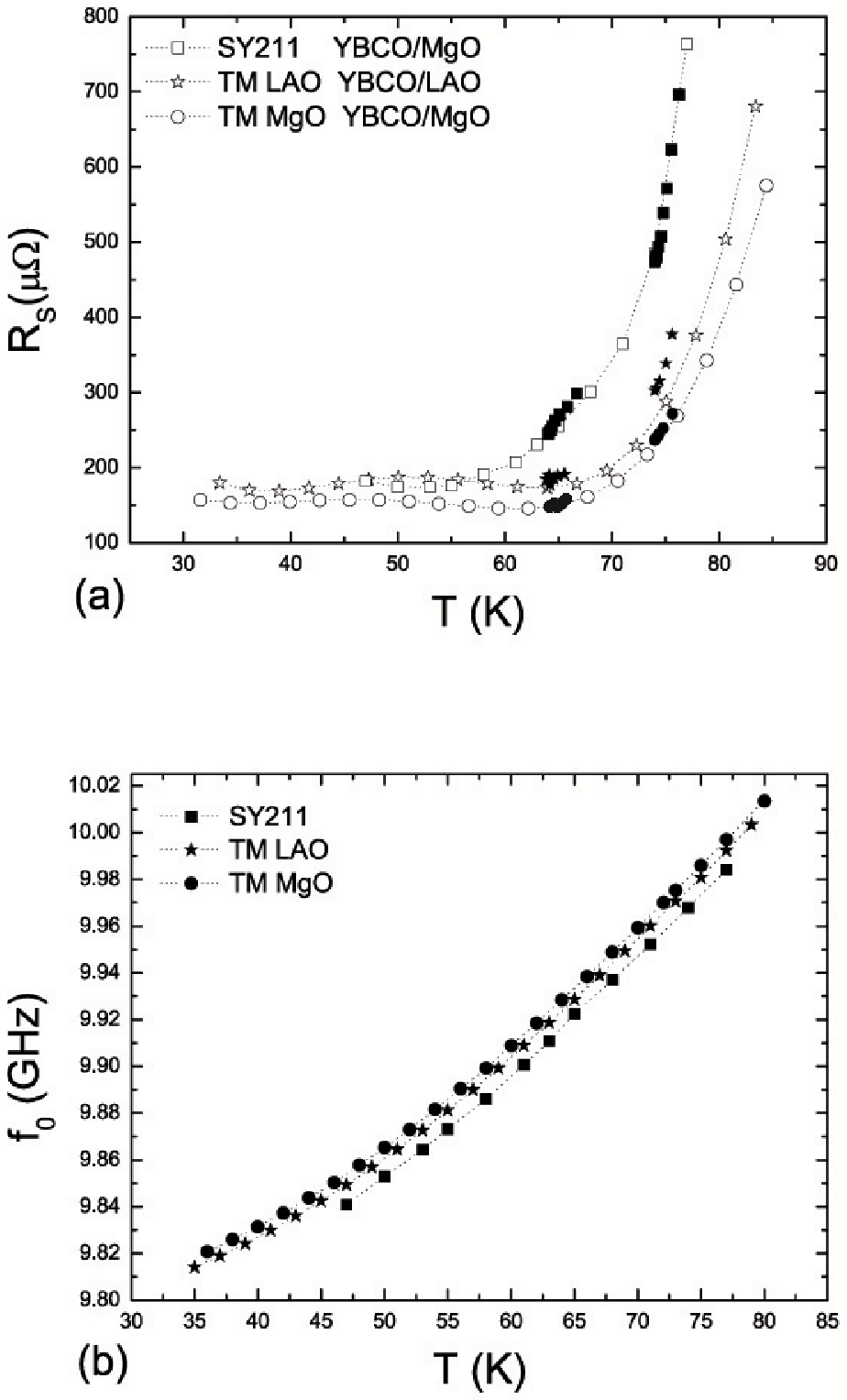} 
\begin{center}  Kermorvant et al. Fig. 5 \end{center}
\end{figure}

\pagebreak

\begin{figure}[h!]
\includegraphics[width=8cm,keepaspectratio]{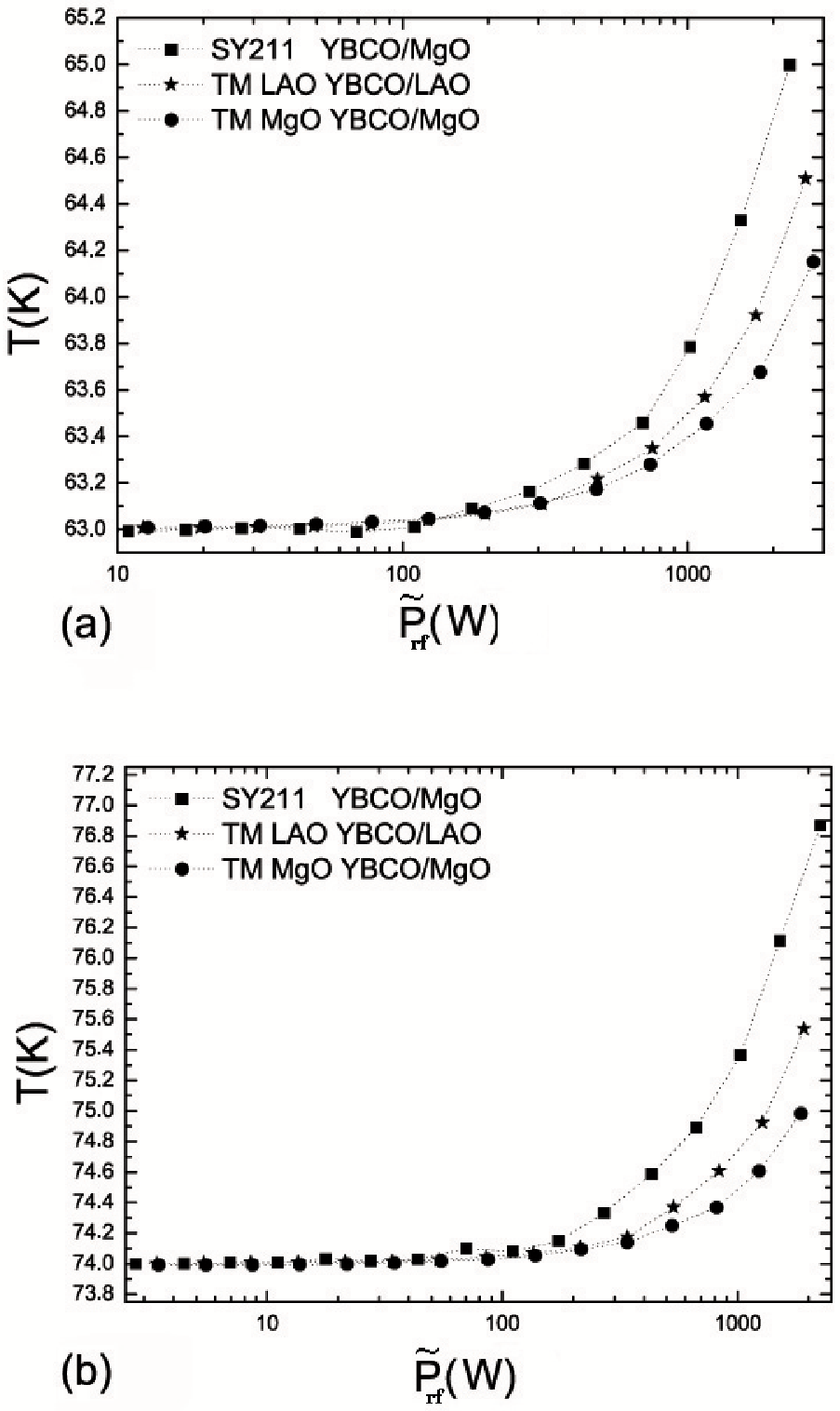} 
\begin{center}  Kermorvant et al. Fig. 6 \end{center}

\end{figure}

\pagebreak

\begin{figure}[h!]
\includegraphics[width=8cm,keepaspectratio]{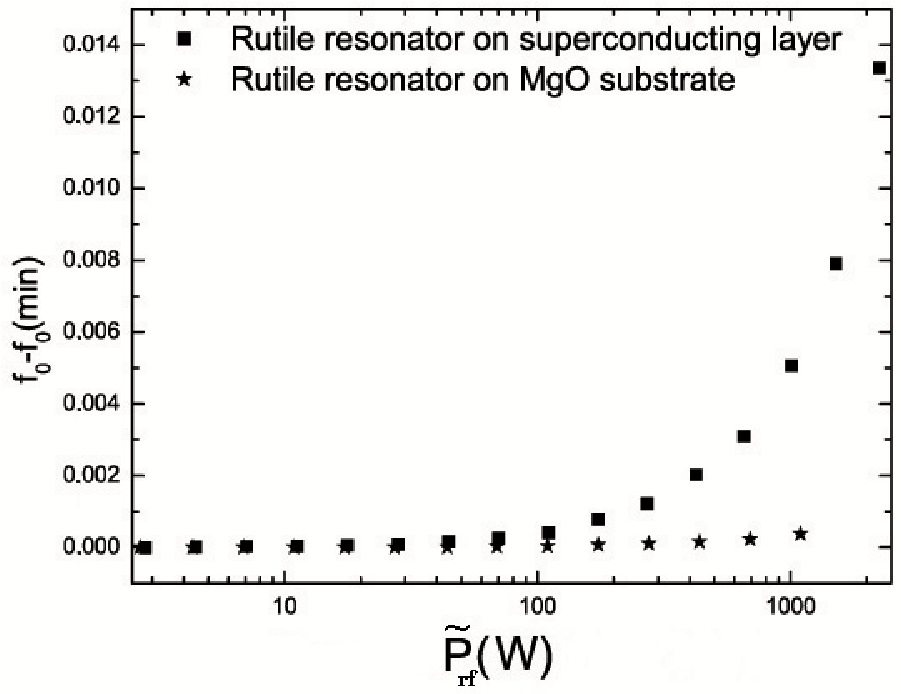} 
\begin{center}  Kermorvant et al. Fig. 7 \end{center}
\end{figure}
\pagebreak

\end{document}